\documentclass[12pt,a4paper]{article}

\usepackage{amssymb,amsmath,eucal}
\usepackage[dvips]{lscape,graphicx}
\usepackage{cite}

\newcommand{\bc}{\begin{center}}
\newcommand{\ec}{\end{center}}
\newcommand{\bd}{\begin{displaymath}}
\newcommand{\ed}{\end{displaymath}}
\newcommand{\be}{\begin{equation}}
\newcommand{\ee}{\end{equation}}
\newcommand{\ba}{\begin{array}}
\newcommand{\ea}{\end{array}}
\newcommand{\bt}{\begin{tabular}}
\newcommand{\et}{\end{tabular}}

\newcommand{\ct}{\cite}

\newcommand{\lb}{\label}
\newcommand{\bp}{\begin{picture}}
\newcommand{\ep}{\end{picture}}
\newcommand{\bfi}{\begin{figure}}
\begin{document}

\vspace{3cm}

\title{\Large\bf {Generalized Duality Symmetry of Non-Abelian Theories}}
\author{{\bf L.V.Laperashvili}
\footnote{{\bf E-mail}:laper@heron.itep.ru, laper@alf.nbi.dk}\\
\it Institute of Theoretical and Experimental Physics,\\
\it B.Cheremushkinskaya 25, 117218 Moscow, Russia}

\date{}

\maketitle

\thispagestyle{empty}

\newpage
\thispagestyle{empty}
\vspace{10cm}
\begin{abstract}

The quantum Yang--Mills theory describing dual ($\tilde g$) and
non--dual ($g$) charges and revealing the generalized duality
symmetry was developed by analogy with the Zwanziger formalism in QED.

\end{abstract}

\newpage

\pagenumbering{arabic}
\vspace{1cm}
\large
\section{ Introduction}
\vspace{0.51cm}

In the last years gauge theories essentially operate with the
fundamental idea of duality (see, for example, reviews \ct{1} and
references there).

Duality is a symmetry appearing in pure electrodynamics as invariance
of the free Maxwell equations:
\be
 \bf{\bigtriangledown}
\cdot \vec {\bf B} = 0, \quad\quad
 \bf{\bigtriangledown}\times \vec {\bf E} = - \partial_0\vec {\bf B},
                 \lb{1}
\ee
\be
 \bf{\bigtriangledown}\cdot \vec{\bf E} = 0, \quad\quad
 \bf{\bigtriangledown}\times \vec{\bf B} = \partial_0\vec{\bf E},
                 \lb{2}
\ee
under the interchange of electric and magnetic fields:
\be
    \vec {\bf E} \to \vec {\bf B}, \quad\quad
        \vec{\bf B} \to  -\vec{\bf E}.                  \lb{3}
\ee
Letting
\be
      F = \partial\wedge A = - (\partial\wedge B)^{*},        \lb{4}
\ee
\be
      F^{*} = \partial\wedge B = (\partial\wedge A)^{*},        \lb{5}
\ee
it is easy to see that the equations of motion:
\be
    \partial_\lambda F_{\lambda\mu} = 0,                   \lb{6}
\ee
which together with the Bianchi identity:
\be
    \partial_\lambda F_{\lambda\mu}^{*} = 0                \lb{7}
\ee
are equivalent to Eqs.(\ref{1}) and (\ref{2}), show the invariance under
the Hodge star operation on the field tensor:
\be
 F_{\mu\nu}^{*} = \frac 12 \epsilon_{\mu\nu\rho\sigma} F_{\rho\sigma}.
                                                  \lb{8}
\ee
This Hodge star duality, having a long history \ct{2a}-\ct{4},
does not hold in general for non--Abelian theories.
In Abelian theory Maxwell's equation (\ref{6})
is equivalent to the Bianchi identity for the dual field $F_{\mu\nu}^{*}$,
which guarantees the existence of a "dual potential" (see $B_{\mu}$ in
Eq.(\ref{5})).

In the non--Abelian Yang--Mills theory, one usually starts with a gauge
field $F_{\mu\nu}(x)$ derivable from a potential $A_{\mu}(x)$:
\be
   F_{\mu\nu} = \partial_{\nu}A_{\mu}(x) - \partial_{\mu}A_{\nu}(x) +
       ig [ A_{\mu}(x), A_{\nu}(x) ].           \lb{9}
\ee
Considering (for simplicity of presentation) only gauge group with Lie
algebra SU(N), we have :
\be
      A_{\mu}(x) = t^j A_{\mu}^j(x),\quad \quad
           j = 1,...,N^2 - 1,              \lb{10}
\ee
where $t^j$ are the generators of SU(N) group.
Equations of motion obtained by extremizing the corresponding action
with respect to $A_{\mu}(x)$ gives:
\be
         D_{\nu}F^{\mu\nu}(x) = 0,                    \lb{11}
\ee
where $D_{\nu}$ is the covariant derivative defined as
\be
             D_{\mu} = \partial_{\mu} - ig[ A_{\mu}(x),\quad ].     \lb{12}
\ee
The analogy to electromagnetism is still rather close. But Yang--Mills
equation (\ref{11}) does not imply in general the existence of a
potential for the corresponding dual field $F_{\mu\nu}^{*}$. This
Yang--Mills equation itself can no longer be interpreted
as the Bianchi identity for $F_{\mu\nu}^{*}(x)$, nor does it imply the
existence of a "dual potential" ${\tilde A}_{\mu}(x)$ satisfying
\be
    F_{\mu\nu}^{*}(x) \stackrel{?}{=} \partial_{\nu}{\tilde A}_{\mu}(x) -
\partial_{\mu}{\tilde A}_{\nu}(x) + i{\tilde g} [ {\tilde A}_{\mu}(x),
{\tilde A}_{\nu}(x) ],             \lb{13}
\ee
in parallel to (\ref{9}). This result means that the dual symmetry
of the Yang--Mills theory under the Hodge star operation does not hold.
So one has to seek a more general form of duality for non--Abelian theories
than the Hodge star operation on the field tensor.

\section{The Generalized Polyakov Variables}

It was shown in Refs.\ct{2},\ct{3} that the classical Yang--Mills theory
is symmetric under a generalized dual transform which reduces to the
well--known electric--magnetic duality in the Abelian case. This dual
transform
is formulated in Refs.\ct{2}-\ct{4} in terms of loop variables similar to
those introduced by A.M.Polyakov \ct{5}.

Starting with the Dirac phase factor \ct{6} :
\be
   \Phi [\xi ] = P_s \exp [ ig\int_0^{2\pi} ds A_{\mu}\bigl(\xi(s)\bigl)
              \dot {\xi}^{\mu}(s) ]
                                            \lb{14}
\ee
for a parametrized closed loop $\xi (s)$, $s = 0 \to {2\pi}$, one can
define the Polyakov variables \ct{5} :
\be
  F_{\mu}[\xi|s] = \frac ig {\Phi}^{-1}[\xi] \frac{\delta \Phi[\xi]}
                   {\delta \xi^{\mu}(s)} .           \lb{15}
\ee
The duality transformation proposed in Refs.\ct{2},\ct{3} operates
on the following variables:
\be
    E_{\mu}[\xi|s] = \Phi_{\xi}(s,0)F_{\mu}[\xi|s]{\Phi_{\xi}}^{-1}(s,0) ,
                                                       \lb{16}
\ee
where
\be
   \Phi_{\xi}(s_1, s_2) = P_s\exp [ ig\int_{s_1}^{s_2}ds A_{\mu}
\bigl(\xi(s)\bigl) \dot {\xi}^{\mu}(s) ] .                    \lb{17}
\ee
Therefore $\Phi [\xi]\equiv \Phi_{\xi}(0, 2\pi)$.

With aim to understand the difference between the quantitites
$F_{\mu}[\xi|s]$ and $E_{\mu}[\xi|s]$, it is convinient to give
some explanations.

The loop derivative in Eq.(\ref{15}) is defined as
\be
  \frac{\delta \Phi[\xi ]}{\delta \xi^{\mu}(s)} = \lim_{\Delta \to 0}
         \frac {\Phi [\xi'] - \Phi [\xi]}{\Delta},
                                                          \lb{18}
\ee
where
\be
    {\xi'}^{\lambda} = {\xi}^{\lambda}(s') +
            \Delta \delta^{\lambda}_{\mu}\delta (s - s').      \lb{19}
\ee
The $\delta$--function $\delta (s-s')$ is
a bump function centred at s with width $\epsilon = s_{+} - s_{-}$ (see
\ct{2}).

In contrast to $F_{\mu}[\xi|s]$, the quantity $E_{\mu}[\xi|s]$
depends only on a "segment" of the loop $\xi^{\mu}(s)$ from $s_{-}$
to $s_{+}$.

The regularization of $\delta$--function is necessary
for the definition of loop derivatives used in this theory.

The quantities $E_{\mu}[\xi|s]$ constrained by the condition:
\be
    \frac{\delta E_{\mu}[\xi|s]}{\delta \xi^{\nu}} -
    \frac{\delta E_{\nu}[\xi|s]}{\delta \xi^{\mu}} = 0       \lb{20}
\ee
constitute a set of the curl--free variables valid for the description
of Yang--Mills theories revealing  properties of the generalized dual
symmetry. Here it is necessary to note that, in contrast to the
Polyakov variables $F_{\mu}[\xi|s]$, the variables $E_{\mu}[\xi|s]$
are gauge dependent quantities. But in spite of this unconvinient
property, the variables $E_{\mu}[\xi|s]$ are more useful for studying
the generalized duality.

The authors of Refs.\ct{2},\ct{3} consider also the dual variables
${\tilde E}_{\mu}[\eta|t]$
defined by the following relation:
$$
  {\omega}^{-1}\bigl(\eta(t)\bigl) {\tilde E}_{\mu}[\eta|t]\omega\bigl(
     \eta(t)\bigl) =
$$
\be
- \frac 2K {\epsilon}_{\mu\nu\rho\sigma}
     \dot {\eta}^{\nu}(t)\int \delta{\xi}ds E^{\rho}[\xi|s]
     \dot {\xi}^{\sigma}(s)
     \dot {\xi}^{-2}(s)\delta\bigl(\xi(s) - \eta(t)\bigl) ,
                                                     \lb{21}
\ee
where $K$ is an (infinite) normalization constant:
\be
        K = \int_0^{2\pi}ds{\Pi}_{s'\neq s}d^4\xi(s').       \lb{22}
\ee
The integral in Eq.(21) is over all loops and over all points of each loop,
and the factor $\omega(x)$ is just a rotational matrix allowing for the
change of local frames between the two sets of variables.

As it was shown in Refs.\ct{2},\ct{3}, the expression (\ref{21})
reduces to the Hodge star operation in the Abelian case,
but for a non--Abelian theory they are in general different.

The usual Yang--Mills action
\be
 S_0 = - \frac 1{16\pi}\int d^4x Tr(F_{\mu\nu}F^{\mu\nu})        \lb{23}
\ee
can be expressed in terms of the new variables :
\be
S_0 = - \frac 1{4\pi K}\int \delta\xi ds Tr(E_{\mu}E^{\mu})\dot {\xi}^{-2} ,
                                                          \lb{24}
\ee
or dually :
\be
{\tilde S}_0 = S_0 = - \frac 1{4\pi K}\int \delta \eta dt
Tr({\tilde E}_{\mu}{\tilde E}^{\mu})\dot {\eta}^{-2} .      \lb{25}
\ee
The quantities ${\tilde E}_{\mu}[\xi|s]$ are also described by
Eqs.(\ref{14})-(\ref{17}) with the following replacements:
$$
   g \to \tilde g, \quad s \to t, \quad \xi^{\mu}(s) \to \eta^{\mu}(t),
   \quad
   \Phi[\xi] \to \tilde \Phi [\eta], \quad \Phi_{\xi}(s_1,s_2) \to
   {\tilde \Phi}_{\eta}(\eta_1,\eta_2),
$$
\be
   F_{\mu}[\xi|s] \to {\tilde F}_{\mu}[\xi|s], \quad
   E_{\mu}[\xi|s] \to {\tilde E}_{\mu}[\xi|s].
                     \lb{26}
\ee
Summarizing the results of Refs.\ct{2},\ct{3}, we can formulate such
items :

1. $E_{\mu}[\xi|s]$ is derivable from a local potential $A_{\mu}(x)$ if
and only if the (loop) curl of $E_{\mu}[\xi|s]$ vanishes
(see Eq.(\ref{20})).

2. The constrained action contains the Lagrange multipliers
$W^{\mu\nu}[\xi|s]$:
\be
    S = S_0 + \int \delta \xi ds Tr\bigl( W^{\mu\nu}[\xi|s]
        (\delta E_{\mu}[\xi|s]/\delta{\xi}^{\nu}(s) -
        \delta E_{\nu}[\xi|s]/\delta {\xi}^{\mu}(s))\bigl)
                                                       \lb{27}
\ee
and implies
\be
      \frac{\delta E_{\mu}}{\delta {\xi}^{\mu}(s)} = 0 ,     \lb{28}
\ee
which is equivalent to the Yang--Mills equation (\ref{11}).

3. The Eq.(\ref{28}) is equivalent to the dual variables ${\tilde E}_{\mu}
[\eta|t]$ being curl--free, which (according to the point 1) is equivalent
to the existence of a local potential ${\tilde A}_{\mu}(x)$ :
\be
{\tilde F}_{\mu\nu}(x) = \partial_{\nu}{\tilde A}_{\mu}(x) -
                         \partial_{\mu}{\tilde A}_{\nu}(x) +
              i{\tilde g} [{\tilde A}_{\mu}(x), {\tilde A}_{\nu}(x)],
                                                    \lb{29}
\ee
obeying the "dual Yang--Mills equation" :
\be
{\tilde D}^{\nu}{\tilde F}_{\mu\nu}(x) = 0 ,      \lb{30}
\ee
where
\be
  {\tilde D}_{\nu} = \partial_{\nu} - i{\tilde g} [{\tilde A}_{\nu}(x),
                                \quad ]                \lb{31}
\ee
is the "dual covariant derivative".

4. The local dual potential ${\tilde A}_{\mu}(x)$ can be expressed in terms
of the Lagrange multipliers $W^{\mu\nu}$ (see \ct{2},\ct{3}) as:
\be
   {\tilde A}_{\mu}(x) = \int \delta \xi ds \epsilon_{\mu\nu\rho\sigma}
    \omega\bigl(\xi(s)\bigl)W^{\rho\sigma}[\xi|s]
    \omega^{-1}\bigl(\xi(s)\bigl)\dot {\xi}^{\nu}(s)\dot {\xi}^{-2}(s)
    \delta(\xi(s) - x) .
                                     \lb{32}
\ee

5. The following charge quantization condition exists in theory \ct{2}:
\be
             g\tilde g = 4\pi n, \quad n\in Z.       \lb{32a}
\ee

\section{The Zwanziger--type Action for non--Abelian Theories}

Following the idea of Zwanziger \ct{7},\ct{8} (see also \ct{9},\ct{10})
to describe symmetrically dual and non-dual Abelian fields covariantly
interacting with magnetic and electric currents (respectively), we
suggest to consider the generalized Zwanziger formalism for non-
Abelian theories. The action of such theories is based on the Chan--Tsou
generalized dual symmetry and has the following form:
$$
   S = - \frac 2K \int \delta{\xi}ds \{ Tr\bigl(E^{\mu}[\xi|s]
     E_{\mu}[\xi|s]\bigl) + Tr \bigl({\tilde E}^{\mu}[\xi|s]
     {\tilde E}_{\mu}[\xi|s]\bigl)
$$
\be
    + i Tr\bigl(E^{\mu}[\xi|s]{\tilde E}_{\mu}^{(d)}[\xi|s]\bigl) +
     i Tr\bigl({\tilde E}^{\mu}[\xi|s]
     E_{\mu}^{(d)}[\xi|s]\bigl)\} {\dot\xi}^{-2}(s) + S_{gf},
                        \lb{33}
\ee
where $S_{gf}$ is the gauge--fixing action. The choice
\be
    S_{gf} = \frac 2K \int \delta\xi ds [ M_A^2 {(\dot{\xi}\cdot A)}^2 +
             M_B^2{({\dot\xi}\cdot \tilde A)}^2 ]{\dot\xi}^{-2}
                        \lb{34}
\ee
exludes ghosts in the theory.

In Eq.(\ref{33}) we have used the generalized dual operations:
$$
  E_{\mu}^{(d)}[\xi|s] =
$$
\be
- \frac 2K \epsilon_{\mu\nu\rho\sigma}{\dot\xi}^{\nu}
  \int \delta{\eta}dt \omega \bigl(\eta(t)\bigl)E^{\rho}[\eta|t]
  {\omega}^{-1}\bigl(\eta(t)\bigl){\dot\eta}^{\sigma}(t){\dot\eta}^{-2}
  \delta \bigl(\eta(t) - \xi(s)\bigl),      \lb{35a}
\ee
$$
 {\tilde E}_{\mu}^{(d)}[\xi|s] =
$$
\be
\frac 2K \epsilon_{\mu\nu\rho\sigma}
 {\dot\xi}^{\nu} \int \delta{\eta}dt \omega^{-1}\bigl(\eta(t)\bigl)
 {\tilde E}^{\rho}[\eta|t]\omega \bigl(\eta(t)\bigl){\dot\eta}^{\sigma}(t)
 {\dot\eta}^{-2}\delta \bigl(\eta(t) - \xi(s)\bigl).
                    \lb{35}
\ee
From action (\ref{33}) we have the following equations of motions:
\be
        \frac{\delta E_{\mu}[\xi|s]}{\delta \xi^{\mu}(s)} = 0, \lb{35A}
\ee
\be
 \frac{\delta \tilde E_{\mu}[\xi|s]}{\delta \xi^{\mu}(s)} = 0. \lb{35B}
\ee
In the Abelian case, the following relations are easily obtained:
\be
     E_{\mu}[\xi|s] =
           F_{\mu\nu}\bigl(\xi(s)\bigl){\dot\xi}^{\nu}(s),
                            \lb{36}
\ee
\be
 {\tilde E}_{\mu}[\xi|s] = F_{\mu\nu}^{*}\bigl(\xi(s)\bigl){\dot\xi}^{\nu}(s),
                     \lb{37}
\ee
\be
     E_{\mu}^{(d)}[\xi|s] =
   - \frac 12 \epsilon_{\mu\nu\rho\sigma}{\dot\xi}_{\nu}F_{\rho\sigma} =
          - {\dot\xi}\cdot {(\partial \wedge A)}^{*}
                               \lb{38}
\ee
and
\be
   {\tilde E}_{\mu}^{(d)}[\xi|s] =
 \frac 12 \epsilon_{\mu\nu\rho\sigma}{\dot\xi}_{\nu}{\tilde F}_{\rho\sigma} =
          {\dot\xi}\cdot {(\partial \wedge \tilde A)}^{*}.
                              \lb{39}
\ee
Introducing a unit vector:
\be
    n_s^{\mu}\equiv n^{\mu}(s) = {\dot\xi}^{\mu}/\sqrt{{\dot\xi}^2},
                       \lb{40}
\ee
tangential to the loop $\xi^{\mu}=\xi^{\mu}(s)$, we obtain the following
action for Abelian fields:
$$
   S = - \frac 2K \int \delta{\xi}ds
      \{{\bigl(n_s\cdot [\partial \wedge A]\bigl)}^2 +
      {\bigl(n_s\cdot [\partial \wedge \tilde A]\bigl)}^2 +
$$
\be
      i\bigl(n_s\cdot [\partial \wedge A]\bigl)
      \bigl(n_s\cdot {[\partial \wedge \tilde A]}^{*}\bigl) -
      i\bigl(n_s\cdot [\partial \wedge \tilde A]\bigl)
      \bigl(n_s\cdot {[\partial \wedge A]}^{*}\bigl)\},
                               \lb{41}
\ee
where
\be
   (n\cdot [A\wedge B]) = n_{\nu}(A_{\mu}B_{\nu} - A_{\nu}B_{\mu}).
                          \lb{42}
\ee
The Zwanziger action for Abelian fields \ct{7} follows immediately from the
action (\ref{41}) if we choose $n_{\mu}$ as a direction of a "frozen string"
imagined as an unclosed loop of the fixed direction in the 4--space:
$$
   S = - \frac 12 \int d^4x
      \{{\bigl(n\cdot [\partial \wedge A]\bigl)}^2 +
      {\bigl(n\cdot [\partial \wedge \tilde A]\bigl)}^2 +
$$
\be
      i\bigl(n\cdot [\partial \wedge A]\bigl)
      \bigl(n\cdot {[\partial \wedge \tilde A]}^{*}\bigl) -
      i\bigl(n\cdot [\partial \wedge \tilde A]\bigl)
      \bigl(n\cdot {[\partial \wedge A]}^{*}\bigl)\}.
                               \lb{43}
\ee

Let us return to the non-Abelian theories. In the light of the
regularization procedure considered in this paper, we have only
the following relations:
\be
    \lim\limits_{\epsilon \to 0}E_{\mu}[\xi|s]
                      = F_{\mu\nu}{\dot\xi}^{\nu}(s),   \lb{44}
\ee
\be
    \lim\limits_{\epsilon \to 0}{\tilde E}_{\mu}[\xi|s]
               = {\tilde F}_{\mu\nu}{\dot\xi}^{\nu}(s).
                                                       \lb{45}
\ee
Now
$$
    \lim\limits_{\epsilon \to 0}E^{(d)}_{\mu}[\xi|s] \neq
   - \frac 12 \epsilon_{\mu\nu\rho\sigma}{\dot\xi}_{\nu}F_{\rho\sigma} ,
$$
\be
    \lim\limits_{\epsilon \to 0}{\tilde E}^{(d)}_{\mu}[\xi|s] \neq
 \frac 12 \epsilon_{\mu\nu\rho\sigma}{\dot\xi}_{\nu}{\tilde F}_{\rho\sigma},
                         \lb{46}
\ee
what means that for non-Abelian theories the reduction to the Hodge
star relation does not go through.

It is obvious now that there exists the second local gauge symmetry
for non--Abelian theories, which can be denoted as $\widetilde {SU(N)}$
to distinguish it from the initial local symmetry of $A_{\mu}$.
As a result, we deal with a doubling of the gauge symmetry from
SU(N) to
\be
             SU(N)\times \widetilde{SU(N)}       \lb{47}
\ee
without extra degrees of freedom.

\section{Conclusions}

It was shown in the present paper that the Zwanziger--type action
can be constructed for non--Abelian theories revealing the generalized
duality symmetry. In the Abelian limit this action corresponds to the
Zwanziger formalism for quantum electro--magneto dynamics (QEMD).
It was emphasized that although
the generalized duality transformation is rather complicated,
it is explicit in terms of the Polyakov loop space variables type.

\vspace{0.5cm}

ACKNOWLEDGMENTS:

The author whould like to express special thanks to Holger Bech Nielsen
for useful discussions, and to D.A.Ryzhikh for help.

I am indebted to the Niels Bohr Institute
for its hospitality and financial support.

\end{document}